# Study of decoherence of a superposition of macroscopic quantum states by means the consideration of a multimode state of a Schrödinger cat

D.V. Fastovets[*], Yu.I. Bogdanov, N.A. Bogdanova, V.F. Lukichev
Valiev Institute of Physics and Technology of Russian Academy of Sciences, Russia, Moscow

**ABSTRACT**

Quantum Schrodinger cat states are of great interest in quantum communications and quantum optics. These states are used in various scientific fields such as quantum computing, quantum error correction and high-precision measurements. The analysis of the Schrodinger cat states coherence is an important task for their complete practical application. Our developed approach makes it possible to estimate the coherence of the quantum Schrodinger cat state of arbitrary dimension, as well as to find the interference visibility of the state – an important optical characteristic. The obtained simple quantitative relationship between coherence and the Schmidt number, as well as the developed approach of reducing the multidimensional quantum cat state to a two-mode analog allow us to analyze macroscopic states formed by a large number of modes. Several explicit formulas for the reduced states that arise after measuring of some modes of the considered multimode system are obtained. The research results have significant application and can be used in the development of high-dimensional quantum information processing systems.

**Keywords:** quantum computation, quantum optics, interference, entanglement, decoherence

## 1. INTRODUCTION

Interference of quantum states is an important aspect of quantum computing [1]. This effect is observed in various systems: a diffraction grating, biphoton fields, double-beam electronic interferometers [2], etc. It is important to note the interference effects manifested in quantum states of the Schrödinger cat, which are a superposition of coherent states with different phase [3]. These states are actively used in quantum optics and optical technologies [4-6]. In addition, these states are used in other scientific fields, for example, computations in continuous variables [7-9], quantum error correction codes [10,11] and precision measurements [12,13]. Such a widespread usage of quantum Schrödinger cat states makes them a universal practical tool.

However, the generation of multimode quantum Schrödinger cat states is an extremely difficult problem. Thus, the direct transform of the light coherent state into the Schrödinger cat state requires the creation of an environment with significant nonlinearity [14]. Scaling and generating multimode Schrödinger cat states is a rather difficult task also due to the presence of decoherence. However, most applications require that the coherent Schrödinger cat states have sufficiently high values of the average number of photons and number of modes [8,15]. The problem of generating multimode Schrödinger cat states with a high average number of photons is important and relevant both from the point of view of fundamental science and from the point of view of applied interest in metrology, quantum computing algorithms, etc.

## 2. MATHEMATICAL APPARATUS FOR THE ANALYSIS OF INTERFERING ALTERNATIVES

Let consider a bipartite system consisting of subsystems $A$ and $B$. Let there be two interfering alternatives $|\varphi_1\rangle$ and $|\varphi_2\rangle$ of the first subsystem entangled with the corresponding alternatives states of the second subsystem: $|\psi_1\rangle$ and $|\psi_2\rangle$ (we assume that all introduced states are normalized to unity):

*fast93@mail.ru

$$|\psi\rangle = \frac{1}{\sqrt{2 + q_1 q_2 + q_1^* q_2^*}} \left( |\varphi_1, \psi_1\rangle + |\varphi_2, \psi_2\rangle \right), \tag{1}$$

where $q_1 = \langle \varphi_1 | \varphi_2 \rangle$ - probability amplitude of detecting alternative $|\varphi_1\rangle$, provided that alternative $|\varphi_2\rangle$ was prepared. Similarly $q_2 = \langle \psi_1 | \psi_2 \rangle$ - probability amplitude of coincidence of the second subsystem.

This entangled system (1) can be reduced to an effective two-qubit system, regardless of the complexity of the interfering states. In this case, the first qubit specifies the interfering alternatives of subsystem $A$, and the second – subsystem $B$. Using the orthogonalization procedure, it is easy to obtain the basis states of the considered qubits.

$$|0\rangle_1 = |\varphi_1\rangle, \quad |1\rangle_1 = \frac{1}{\sqrt{1-|q_1|^2}} \left( |\varphi_2\rangle - q_1 |\varphi_1\rangle \right);$$

$$|0\rangle_2 = |\psi_1\rangle, \quad |1\rangle_2 = \frac{1}{\sqrt{1-|q_2|^2}} \left( |\psi_2\rangle - q_2 |\psi_1\rangle \right). \tag{2}$$

As a result, the initial state (1) can be represented as

$$|\psi\rangle = \frac{1}{\sqrt{2 + q_1 q_2 + q_1^* q_2^*}} \left( c_{00} |00\rangle + c_{01} |01\rangle + c_{10} |10\rangle + c_{11} |11\rangle \right), \tag{3}$$

where $c_{00} = 1 + q_1 q_2$, $c_{01} = q_1 \sqrt{1-|q_2|^2}$, $c_{10} = q_2 \sqrt{1-|q_1|^2}$, а $c_{11} = \sqrt{(1-|q_1|^2)(1-|q_2|^2)}$.

To analyze correlation and other characteristics of state (3) it is convenient to use the Schmidt decomposition [16]. This decomposition allows us to represent an arbitrary pure two-particle (consisting of subsystems $A$ and $B$) state $|\psi\rangle$ in the form [1]:

$$|\psi\rangle = \sum_{k=1}^{s} \sqrt{\lambda_k} |\psi_k^A\rangle |\psi_k^B\rangle, \tag{4}$$

where $\lambda_k$ - Schmidt weights coefficients sorted in descending (non-increasing) order, $|\psi_k^A\rangle$ and $|\psi_k^B\rangle$ – corresponding Schmidt modes of subsystems $A$ and $B$, $s$ – dimension of the smallest subsystem, $s = \min(\dim A, \dim B)$. Based on the set of Schmidt coefficients, we can introduce characteristic that describes the effective number of modes - the Schmidt number:

$$K = \frac{1}{\sum_k \lambda_k^2}. \tag{5}$$

The Schmidt number lies on the segment $1 \leq K \leq s$ and allows us to estimate the degree of relationship between two subsystems: $K = 1$ corresponds to only one nonzero term in decomposition (4) and, therefore, the absence of correlation and quantum entanglement; $K = s$ corresponds to maximal correlation and entanglement between subsystems.

In the case of a two-qubit representation of the interfering alternatives state (3), it is easy to show that for the Schmidt number the following formula will be valid [17]:

$$K = \frac{1}{1 - 2\Delta}, \tag{6}$$

where $\Delta = |c_{00} c_{11} - c_{01} c_{10}|^2 = \frac{(1-|q_1|^2)(1-|q_2|^2)}{(2 + q_1 q_2 + q_1^* q_2^*)^2}$. Using this parameter, we can also obtain formulas for the Schmidt coefficients:

$$\lambda_k = \frac{1}{2}\left(1+(-1)^k \sqrt{1-4\Delta}\right), \tag{7}$$

where $k=0,1$ - coefficient index, limited by the dimension of the minimal subsystem (qubit).

The interference effect, in the theory of optical phenomena, is closely related to the concept of the interference visibility $V$ [18]. This parameter characterizes the intensity modulation by interference fringes and lies on the segment $0 \leq V \leq 1$. A visibility value of zero corresponds to a uniformly illuminated screen and therefore no interference fringes. A visibility value of one indicates the contrast and clarity of the interference fringes on the screen. The interference visibility (for narrow slits) is determined in classical optics by the formula [18]:

$$V = \frac{I_{max} - I_{min}}{I_{max} + I_{min}}. \tag{8}$$

Here $I_{max}$ and $I_{min}$ - maximum and minimum intensity of the recorded optical signal. In the terms of the Schmidt decomposition, the weight of the base (zero) mode $\lambda_0$ interpreting as useful signal $I_{max}$, and the weight of the first mode $\lambda_1$ acts as noise $I_{min}$, therefore we obtain the following relation between visibility, Schmidt number and the parameter $\Delta$ [19]:

$$V = \lambda_0 - \lambda_1 = \sqrt{\frac{2-K}{K}} = \sqrt{1-4\Delta}. \tag{9}$$

An important case of using the developed method for analyzing interfering alternatives corresponds to the situation when the environment of the system under consideration $A$ acts as a subsystem $B$. Consideration of entanglement between the quantum system and its environment sheds light on the nature of the interfering quantum states coherence.

In this case, for clearly distinguishable alternatives, when $q_1 = \langle \varphi_1 | \varphi_2 \rangle$ the following simple relationship between the visibility and coherence of the environment states is valid:

$$V = |q_2| = |\langle \psi_1 | \psi_2 \rangle|, \tag{10}$$

where states $|\psi_1\rangle$ and $|\psi_2\rangle$ are used as states of environment alternatives. Thus, the dot product $\langle \psi_1 | \psi_2 \rangle$ is a generalization of the classical complex parameter $\gamma$, called the coherence degree of light oscillations [20].

The developed mathematical apparatus can be applied to any system defined by interference of two different alternatives interacting with the environment. In this work, we consider the multimode states of the Schrödinger cat and investigate their interference properties.

## 3. THE TWO-MODE QUANTUM STATE OF THE SCHRÖDINGER CAT

The Schrödinger's cat state is a superposition of coherent states differing in phase by $\pi$. If there are two modes, this state will have the form:

$$|cat_{\alpha\beta}\rangle = \frac{1}{\sqrt{2+2q_\alpha q_\beta}}\left(|\alpha,\beta\rangle + |-\alpha,-\beta\rangle\right), \tag{11}$$

where $q_\alpha = \langle \alpha | -\alpha \rangle = \exp(-2|\alpha|^2)$, $q_\beta = \langle \beta | -\beta \rangle = \exp(-2|\beta|^2)$. These states play an important role in the implementation of various quantum algorithms [21], and also allow to implement the quantum key distribution protocols [22].

For a visual representation of the two-mode state of the Schrödinger cat, it is necessary to use the coordinate and momentum representation of the wave function. The wave function corresponding to state (11), has the following form in the coordinate representation (the insignificant phase factor is skipped):

$$\psi_{\alpha\beta}(x,y) = \frac{\sqrt{2}C_{\alpha\beta}}{\sqrt{\pi}} \exp\left(-\frac{x^2+y^2}{2}\right) \cosh\left((\alpha x + \beta y)\sqrt{2}\right), \tag{12}$$

where $C_{\alpha\beta} = \left(\exp\left(2\bar{\alpha}^2 + 2\bar{\beta}^2\right) + \exp\left(-2\bar{\bar{\alpha}}^2 - 2\bar{\bar{\beta}}^2\right)\right)^{-1/2}$ - normalization factor, where the following notations for the real and imaginary parts of the coherent states amplitudes are introduced: $\bar{\alpha} = \text{Re}(\alpha)$, $\bar{\bar{\alpha}} = \text{Im}(\alpha)$, $\bar{\beta} = \text{Re}(\beta)$, $\bar{\bar{\beta}} = \text{Im}(\beta)$.

To find the momentum representation of the wave function, it is necessary to find the Fourier transform of function (12). Then we obtain:

$$\tilde{\psi}_{\alpha\beta}(p_x, p_y) = \frac{\sqrt{2}\tilde{C}_{\alpha\beta}}{\sqrt{\pi}} \exp\left(-\frac{p_x^2 + p_y^2}{2}\right) \cos\left((\alpha p_x + \beta p_y)\sqrt{2}\right), \qquad (13)$$

where $\tilde{C}_{\alpha\beta} = \left(\exp\left(2\bar{\bar{\alpha}}^2 + 2\bar{\bar{\beta}}^2\right) + \exp\left(-2\bar{\alpha}^2 - 2\bar{\beta}^2\right)\right)^{-1/2}$ - normalization factor. Note that the normalization factors $C_{\alpha\beta}$ and $\tilde{C}_{\alpha\beta}$ are related by the formula $\tilde{C}_{\alpha\beta} = C_{i\alpha,i\beta}$.

Since this system (11) contains interfering alternatives, then, to study the quantum correlation (entanglement) between modes, as well as the phenomena of coherence and interference, we apply the general approaches presented in Section 2. Now it should be assumed that $q_1 = q_\alpha$, and $q_2 = q_\beta$.

To demonstrate how a gradual disappearance of the interference (and a decreasing of coherence) occurs, let us fix the value of the parameter $\alpha$ and vary the value of the parameter $\beta$. Figure 1 shows the corresponding interference plot formed by the marginal momentum distribution $P(p_x) = \int \left|\tilde{\psi}_{\alpha\beta}(p_x, p_y)\right|^2 dp_y$ of the first mode momentum.

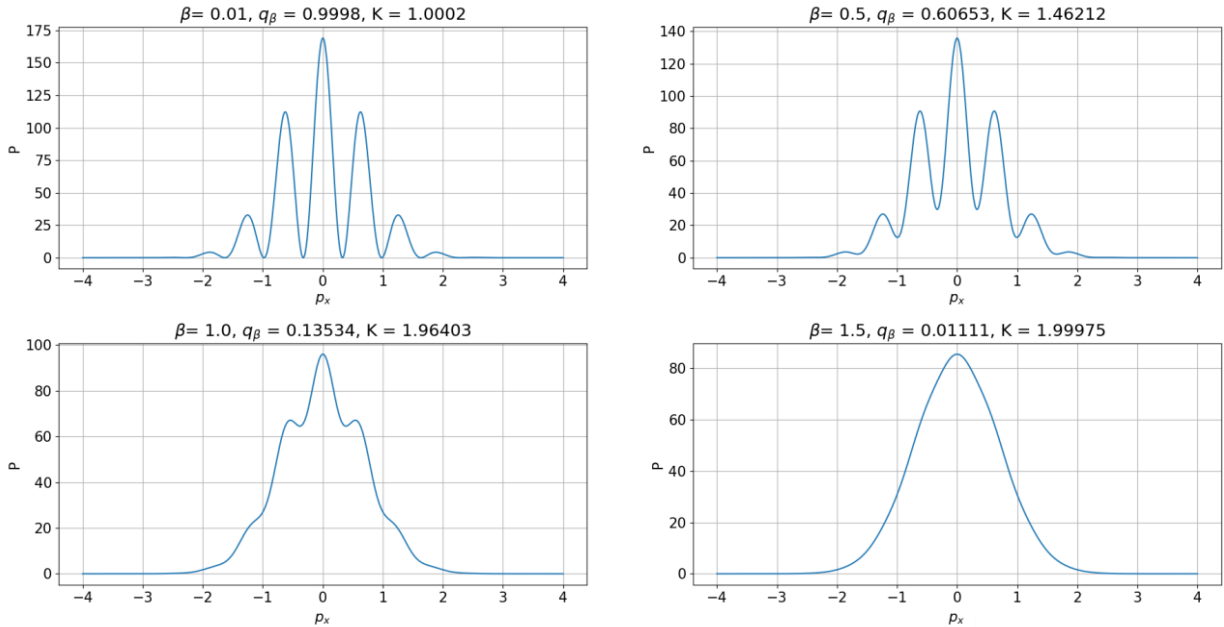

Figure 1. Illustration of the interference disappearance with increasing parameter $\beta$ ($\alpha = 3.4$).

During calculating these plots, the second subsystem (represented by the parameter $\beta$) plays the role of the environment of the main system, represented by the parameter $\alpha$. Thus, the introduced parameter $q_\beta = \langle\beta|-\beta\rangle$ can be interpreting as coherence (or visibility if we take into account that $q_\beta$ is a real, positive number and $q_\alpha = \exp(-2|\alpha|^2)$ is small).

# 4. THE MULTIMODE QUANTUM STATE OF THE SCHRÖDINGER CAT

By analogy with the two-mode state of the Schrödinger cat (11) can be introduced as state consisting of an arbitrary number of modes. The state of the Schrödinger cat, consisting of $n$ modes, is determined by the formula

$$|cat_{\alpha_1\alpha_2...\alpha_n}\rangle = \frac{1}{\sqrt{2+2\prod_{j=1}^{n} q_{\alpha_j}}}\left(|\alpha_1,\alpha_2,...,\alpha_n\rangle + |-\alpha_1,-\alpha_2,...,-\alpha_n\rangle\right), \tag{14}$$

where $q_{\alpha_j} = \langle\alpha_j|-\alpha_j\rangle = \exp\left(-2|\alpha_j|^2\right)$, $j=1,...,n$.

Let us distinguish two subsystems in state (14). Let the first subsystem $A$ be the main one and contain the first $n-m$ state modes. And the second subsystem $B$ consists of the remaining $m$ modes.

$$\left|\underbrace{\alpha_1,...,\alpha_{n-m}}_{A},\overbrace{\alpha_{n-m+1},...,\alpha_n}^{B}\right\rangle + \left|\underbrace{-\alpha_1,...,-\alpha_{n-m}}_{A},\overbrace{-\alpha_{n-m+1},...,-\alpha_n}^{B}\right\rangle.$$

However, the quantum Schrödinger cat state of arbitrary dimension $n > 2$ can be reduced to an effective two-mode Schrödinger cat state [23]. This reduction procedure retains all the correlation and interference characteristics, but greatly simplifies their calculation. The coherence parameters of both subsystems of the effective state are given by the following formulas

$$a = \sqrt{\sum_{j=1}^{n-m}|\alpha_j|^2}, \quad b = \sqrt{\sum_{j=1}^{m}|\alpha_{n-m+j}|^2}. \tag{15}$$

Then the effective quantum state of the Schrödinger cat will have a form similar to the formula (11):

$$|cat_{ab}\rangle = \frac{1}{\sqrt{2+2q_a q_b}}\left(|a,b\rangle + |-a,-b\rangle\right) \tag{16}$$

To visually interpret the multimode state (14) it is necessary to obtain momentum representation of the wave function, which is constructed by analogy with the formula (13):

$$\tilde{\psi}_{\alpha_1\alpha_2...\alpha_n}(p_1,p_2,...,p_n) = \frac{\sqrt{2}\tilde{C}_{\alpha_1\alpha_2...\alpha_n}}{\sqrt[4]{\pi^n}}\exp\left(-\frac{1}{2}\sum_{j=1}^{n}p_j^2\right)\cos\left(\sqrt{2}\sum_{j=1}^{n}\alpha_j p_j\right), \tag{17}$$

where $\tilde{C}_{\alpha_1\alpha_2...\alpha_n} = \left(\exp\left(2\sum_{j=1}^{n}\bar{\alpha}_j^2\right) + \exp\left(-2\sum_{j=1}^{n}\bar{\bar{\alpha}}_j^2\right)\right)^{-1/2}$ - normalization factor, where the following notations for the real and imaginary parts are introduced: $\bar{\alpha}_j = \text{Re}(\alpha_j)$, $\bar{\bar{\alpha}}_j = \text{Im}(\alpha_j)$. Based on this wave function, we can find the probability distribution of momentum variables $p_1,p_2,...,p_n$, using which it is possible to obtain the marginal distribution of momentum variables of the main system by integrating over the variables of the environment system.

$$\tilde{P}_{\alpha_1\alpha_2...\alpha_n}(p_1,p_2,...,p_{n-m}) = \frac{\tilde{C}^2_{\alpha_1\alpha_2...\alpha_n}}{\sqrt{\pi^{n-m}}}\exp\left(-\sum_{j=1}^{n-m}p_j^2\right)\cdot\left[\exp\left(-2\sum_{j=n-m+1}^{n}\bar{\alpha}_j^2\right)\cos\left(2\sqrt{2}\sum_{j=1}^{n-m}\bar{\alpha}_j p_j\right)+\right.$$
$$\left.+\exp\left(2\sum_{j=n-m+1}^{n}\bar{\bar{\alpha}}_j^2\right)\cosh\left(2\sqrt{2}\sum_{j=1}^{n-m}\bar{\bar{\alpha}}_j p_j\right)\right]. \tag{18}$$

For example, let take the quantum state of the Schrödinger cat, which consists of the same $n = 10^5$ modes. Each of these modes is determined by the coherence parameter $\alpha = 0.01$. To analyze the dynamics of interference characteristics, we will gradually increase the number of modes of the environment system $m = 0,1,2,...$. The visibility and Schmidt number can be calculated using the effective two-mode state (16). Figure 2 shows the dynamics of the interference, visibility and Schmidt number with a gradual increase of the number of environment system modes. During computing these plots, the total momentum of the main system was introduced $p = p_1 + ... + p_{n-m}$.

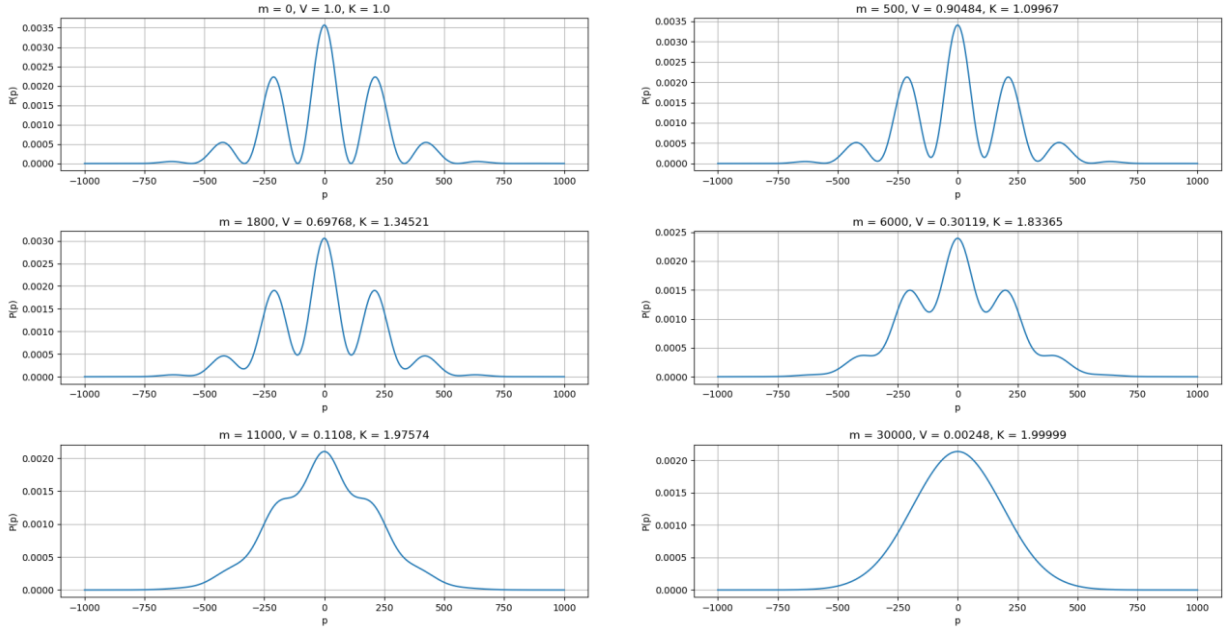

Figure 2. Dynamics of interference's changes with an increasing number of environment modes.

The presented dependences show that the interference visibility rapidly decreases even with the reduction of one photon in the environment modes. For example, using $m = 11000$ modes as an environment corresponds to $1.1$ photon reduction (since in the considered example the average number of photons per one mode is $|\alpha|^2 = 0.0001$).

Here the parameter $V$ is the interference visibility in the case of clearly distinguishable alternatives. It is determined by the number of environment modes and is given by the following simple formula:

$$V = \exp(-2m\alpha^2). \tag{19}$$

The presented analytical formulas allow us to analyze the correlation and optical characteristics for the multidimensional Schrödinger cat state of arbitrary dimension.

## 5. ANALYSIS OF THE COHERENCE OF THE MULTIMODE SCHRÖDINGER CAT STATE

Let consider the coordinate representation of the multimode Schrödinger cat state, which is the Fourier transform of the wave function in the momentum representation (17):

$$\psi_{\alpha_1\alpha_2...\alpha_n}(x_1, x_2, ..., x_n) = \frac{\sqrt{2}C_{\alpha_1\alpha_2...\alpha_n}}{\sqrt[4]{\pi^n}} \exp\left(-\frac{1}{2}\sum_{j=1}^n x_j^2\right) \cosh\left(\sqrt{2}\sum_{j=1}^n \alpha_j x_j\right), \tag{20}$$

where $C_{\alpha_1\alpha_2...\alpha_n} = \left(\exp\left(2\sum_{j=1}^n \bar{\alpha}_j^2\right) + \exp\left(-2\sum_{j=1}^n \bar{\bar{\alpha}}_j^2\right)\right)^{-1/2}$ - normalization factor. Note that the normalization factors of the coordinate and momentum representations are related by a formula similar to the two-mode case $\tilde{C}_{\alpha_1\alpha_2...\alpha_n} = C_{i\alpha_1, i\alpha_2, ..., i\alpha_n}$.

Using (20) we obtain that the probability distribution in the coordinate space $P(x_1, x_2, ..., x_n) = |\psi_{\alpha_1\alpha_2...\alpha_n}(x_1, x_2, ..., x_n)|^2$ has the form:

$$P(x_1, x_2, ..., x_n) = \frac{C^2_{\alpha_1\alpha_2...\alpha_n}}{2\sqrt{\pi^n}} \exp\left(2\sum_{j=1}^{n}|\alpha_j|^2\right)\left[\exp\left(-\sum_{j=1}^{n}(x_j - \bar{\alpha}_j\sqrt{2})^2 + 2\bar{\alpha}_j^2\right) + \right.$$

$$+ \exp\left(-\sum_{j=1}^{n}(x_j - \bar{\bar{\alpha}}_j\sqrt{2})^2 + 2\bar{\bar{\alpha}}_j^2\right) + \exp\left(-\sum_{j=1}^{n}(x_j + \bar{\bar{\alpha}}_j\sqrt{2})^2 + 2\bar{\bar{\alpha}}_j^2\right) + \qquad (21)$$

$$\left. + \exp\left(-\sum_{j=1}^{n}(x_j + \bar{\alpha}_j\sqrt{2})^2 + 2\bar{\bar{\alpha}}_j^2\right)\right].$$

Let consider the case of identical modes $\alpha_1 = \alpha_2 = ... = \alpha_n = \alpha$, assuming, in addition, that $\alpha$ is a real positive number. In this case, formula (21) will have a simpler form:

$$P(x_1, x_2, ..., x_n) = \frac{1}{2\sqrt{\pi^n}} \frac{\exp(2n\alpha^2)}{\exp(2n\alpha^2)+1}\left[\exp\left(-\sum_{j=1}^{n}(x_j - \alpha\sqrt{2})^2\right) + \right.$$

$$+ \exp\left(-\sum_{j=1}^{n}(x_j + \alpha\sqrt{2})^2\right) + \qquad (22)$$

$$\left. + 2\exp\left(-\sum_{j=1}^{n}x_j^2\right)\exp(-2n\alpha^2)\right].$$

If $\alpha$ is a fixed number, then in the limiting case when $n\alpha^2 \gg 1$ we will have $\frac{\exp(2n\alpha^2)}{\exp(2n\alpha^2)+1} \approx 1$, $\exp(-2n\alpha^2) \approx 0$

and as a result, we obtain the following even simpler formula:

$$P(x_1, x_2, ..., x_n) = \frac{1}{2\sqrt{\pi^n}}\left[\exp\left(-\sum_{j=1}^{n}(x_j + \alpha\sqrt{2})^2\right) + \exp\left(-\sum_{j=1}^{n}(x_j - \alpha\sqrt{2})^2\right)\right]. \qquad (23)$$

In this formula, we will associate the first term with the state of the "live cat", and the second - with the state of the "dead cat". At the same time, we conventionally assume that the state of a cat's "health" is determined by a large number of parameters $(x_1, x_2, ..., x_n)$, and positive values of the parameters associated with "health", and negative values - with "painful symptoms".

As in Section 4, we associate the last $m$ state (20) variables with the environment system. To carry out numerical calculations, we select the state parameters $\alpha_1 = \alpha_2 = ... = \alpha_n = \alpha = 0.01$, $n = 10^6$. Using these values, the condition $n\alpha^2 \gg 1$ is true and we can use the formula (23). Let's measure the last mode of the environment subsystem. In this case, we will obtain some coordinate value $x'_n = y_1$. The probability distribution formula (23) in this case takes the form:

$$P(x_1, x_2, ..., x_{n-1}) = \frac{1}{\sqrt{\pi^{n-1}}}\left[p_1^+ \exp\left(-\sum_{j=1}^{n-1}(x_j + \alpha\sqrt{2})^2\right) + p_1^- \exp\left(-\sum_{j=1}^{n-1}(x_j - \alpha\sqrt{2})^2\right)\right], \qquad (24)$$

where $p_1^\pm = \frac{C_1}{2}\exp\left(-(y_1 \pm \alpha\sqrt{2})^2\right)$ - the probabilities of the alternatives "the cat is alive" and "the cat is dead", and $C_1$ - factor providing the normalization condition $p_1^+ + p_1^- = 1$ (this factor depends on the obtained coordinate value $y_1$

). Measurement of the next $(n-1)$-s mode gives the coordinate value $x'_{n-1} = y_2$. Then the formula for the probability distribution (24) is transformed as follows:

$$P(x_1, x_2, ..., x_{n-2}) = \frac{1}{\sqrt{\pi^{n-2}}} \left[ p_2^+ \exp\left(-\sum_{j=1}^{n-2}(x_j + \alpha\sqrt{2})^2\right) + p_2^- \exp\left(-\sum_{j=1}^{n-2}(x_j - \alpha\sqrt{2})^2\right) \right], \quad (25)$$

where $p_2^\pm = \frac{C_2}{2}\exp\left(-\sum_{j=1}^{2}(y_j \pm \alpha\sqrt{2})^2\right)$ - the probabilities of the alternatives "the cat is alive" and "the cat is dead" after two modes measuring, and $C_2$ - factor providing the normalization condition $p_2^+ + p_2^- = 1$. Thus, by measuring all modes of the environment, we will obtain different values of the coordinates $x'_{n-j} = y_j$, $j = 1, 2, ..., m$ which will lead us to the reduced form of the original formula (23):

$$P(x_1, x_2, ..., x_{n-m}) = \frac{1}{\sqrt{\pi^{n-m}}} \left[ p_m^+ \exp\left(-\sum_{j=1}^{n-m}(x_j + \alpha\sqrt{2})^2\right) + p_m^- \exp\left(-\sum_{j=1}^{n-m}(x_j - \alpha\sqrt{2})^2\right) \right], \quad (26)$$

where, by analogy with formulas (24) and (25) we have $p_m^\pm = \frac{C_m}{2}\exp\left(-\sum_{j=1}^{m}(y_j \pm \alpha\sqrt{2})^2\right)$. It is important to note the recurrent connection of the probabilities "the cat is alive" and "the cat is dead" alternatives, which is provided by the sequential measurement of the modes:

$$p_{k+1}^\pm = \frac{C_{k+1}}{C_k} p_k^\pm \exp\left(-(y_{k+1} \pm \alpha\sqrt{2})^2\right), \quad p_0^\pm = \frac{1}{2}, \quad C_0 = 1.$$

Thus, sequential measurements lead to a spontaneous breaking of symmetry between the states "the cat is alive" and "the cat is dead". Indeed, initially, the probabilities of alternatives were equal to each other: $p_0^+ = p_0^- = \frac{1}{2}$. During the coordinates measuring only in one or several modes of the environment, the alternatives remain poorly distinguishable, since in this case $|\alpha|^2 \ll 1$. However, as the number of measured modes increases, the alternatives become more and more distinguishable. This fact can be demonstrated using numerical experiments. Let take several multimode Schrödinger cat states with parameters $\alpha = 0.01$ и $n = 10^6$.

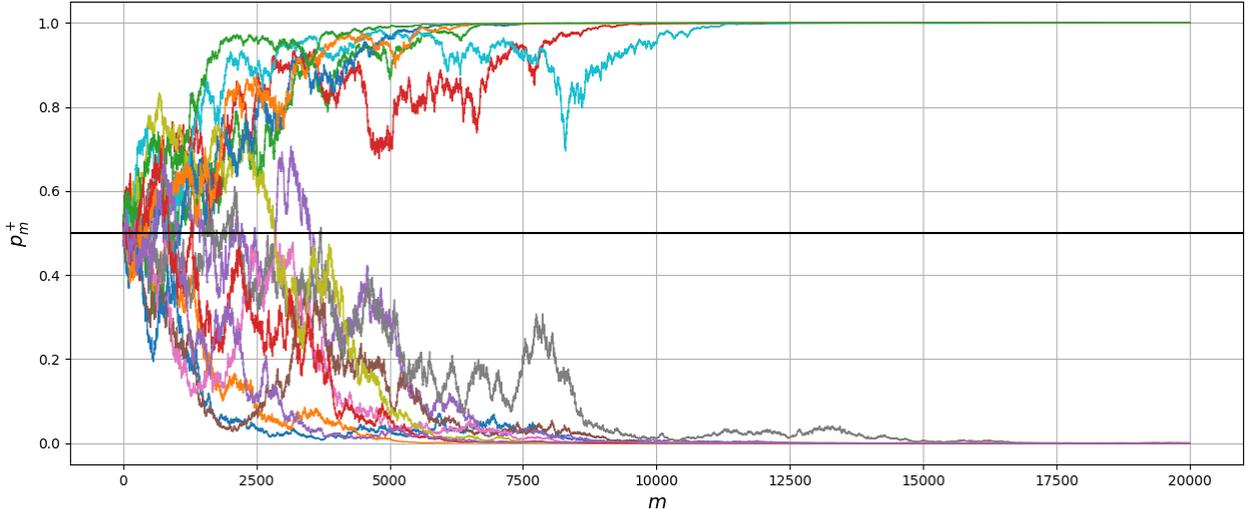

Figure 3. Dependence of the "surviving" probability of the quantum Schrödinger cat state ($\alpha = 0.01$) from the number of reduced modes of the environment. Performed 15 experiments.

In each of these states, we will sequentially measure the environment modes ($m_{max} = 2 \cdot 10^4$ - total number of environment modes). Figure 3 shows the dependence of the state probability "the cat is alive" $p_m^+$ (the probability of "surviving") from the number of measured modes $m$. As we can see from Figure 3, the superposition of the states "the cat is live" and "the cat is dead" cat is almost completely destroyed starting approximately from $m = 15 \cdot 10^3$ (corresponds to a $m\alpha^2 = 1.5$ photon). Therefore, the multimode quantum state of the Schrödinger cat is practically impure, since the reduction of just one photon leads to an almost complete loss of coherence of the initial quantum state.

Let us introduce the "health" parameter ($H$) of the quantum Schrödinger cat state, which will be determined by the formula:

$$H(m) = \ln\left(\frac{p_m^+}{1-p_m^+}\right) = \ln\left(\frac{p_m^+}{p_m^-}\right). \tag{27}$$

This parameter determines the collapse level of the quantum Schrödinger cat state. Obviously, the value $H = 0$ corresponds $p_m^+ = 0.5$, that is, the maximum coherence between the subsystems. Further, if $H \to +\infty$, then $p_m^+ \to 1$ (collapse to "the cat is live" state), and vice versa, if $H \to -\infty$, then $p_m^+ \to 0$ (collapse to "the cat is dead" state). Figure 4 shows plots of the "health" dynamics of the quantum Schrödinger cat states, considered in the Figure 3.

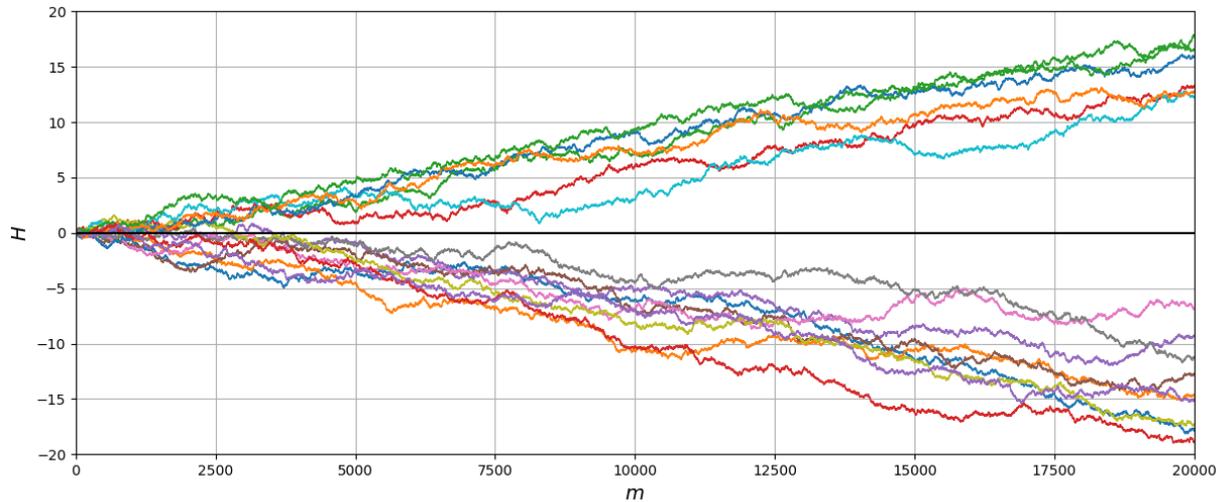

Figure 4. Dependence of the quantum Schrödinger cat state ($\alpha = 0.01$) "health" from the number of reduced modes of the environment. Performed 15 experiments.

It can be shown that the "health" of the quantum Schrödinger cat state is determined by the sum of all measured environmental mode:

$$H(m) = -4\alpha\sqrt{2}\sum_{j=1}^{m} y_j \tag{28}$$

Thus, the obtained characteristic depends on all performed measurements results, as well as the coherence parameter of the quantum state modes.

## 6. CONCLUSIONS

During this work, the mathematical apparatus for analyzing systems with interfering alternatives was developed and investigated. This apparatus allows us to analyze various interference characteristics of systems with arbitrary nature.

The main object of this article is the quantum states of the Schrödinger cat. These states are a system with two interfering alternatives, which allows us to apply to them the tool we have developed. Analytical formulas for calculating the interference visibility in the momentum representation, which determine the degree of coherence of the quantum state between the base system and its environment were obtained.

The obtained results are generalized to the case of multimode Schrödinger cat states. Explicit formulas for the reduced probability distributions arising during measuring some modes of the considered multimode system were obtained. A characteristic that describes the level of collapse of the quantum Schrödinger cat state was introduced. It is shown that the multimode quantum Schrödinger cat state is macroscopically unstable, since the reduction of only one photon leads to an almost complete loss of coherence of the initial quantum state.

## ACKNOWLEDGMENTS

This work was supported by the Ministry of Science and Higher Education of the Russian Federation (program no. FFNN-2022-0016 for the Valiev Institute of Physics and Technology, Russian Academy of Sciences), and by the Foundation for the Advancement of Theoretical Physics and Mathematics BASIS (project no. 20-1-1-34-1)